\documentclass[preprint,showpacs,preprintnumbers,amsmath,amssymb]{revtex4}

\begin{document}
       
\preprint{DESY~08--155\hspace{12.55cm} ISSN 0418--9833}
\preprint{NYU--TH/08/10/31\hspace{14.3cm}}

\boldmath
\title{Quark mixing renormalization effects in the determination of the CKM
parameters $|V_{ij}|$}
\unboldmath

\author{Andrea A. Almasy}
\email{andrea.almasy@desy.de} 
\author{Bernd A. Kniehl}
\email{bernd.kniehl@desy.de}
\affiliation{II. Institut f\"ur Theoretische Physik,
Universit\"at Hamburg, Luruper Chaussee 149, 22761 Hamburg, Germany}
\author{Alberto Sirlin}
\email{alberto.sirlin@nyu.edu}
\affiliation{Department of Physics, New York University,
4 Washington Place, New York, New York 10003, USA}

\date{\today}

\begin{abstract}
We briefly review existing proposals for the renormalization of the
Cabibbo-Kobayashi-Maskawa (CKM) matrix and study the numerical effects of
several of them on the $W$-boson hadronic partial decay widths.
We then use these results to evaluate the relative shifts on the CKM
parameters $|V_{ij}|^2$ induced by the quark mixing renormalization effects,
as well as their scheme dependence.
We also discuss the implications of this analysis for the most precise
unitarity test of the CKM matrix.
\end{abstract}

\maketitle

\section{Introduction} 
 
The renormalizability of the Standard Model (SM) without quark-flavor mixing   
was proved in the early seventies \cite{ref:0}.   
Since the elements of the quark mixing matrices appear as basic parameters in
the bare Lagrangian, they are subject to renormalization, too.   
This is a problem of old vintage \cite{ref:1}, the solution of which was first 
realized for the Cabibbo angle in the SM with two fermion generations in a   
pioneering paper by Marciano and Sirlin \cite{ref:2} in 1975.   
The extension to the Cabibbo-Kobayashi-Maskawa (CKM) quark mixing matrix of  
the three-generation SM was addressed fifteen years later \cite{ref:3}.  
In the subsequent years, interest on the subject increased significantly, and
new renormalization prescriptions were proposed 
\cite{ref:4,ref:6,ref:Y,ref:7,ref:Z,ref:L,ref:DKR,ref:8,ref:12}.   
 
The on-shell (OS) prescription of Ref.~\cite{ref:3} is compact and plausible, 
but the proposed expression for the CKM matrix counterterm, $\delta V$, is 
given in terms of wave-function renormalization constants and is thus gauge 
dependent, as was noticed later \cite{ref:4,ref:6,ref:5}.  
This prescription was employed to study the electroweak effects on the
$B^0$--$\overline{B}^0$ mixing in Ref.~\cite{Gambino:1998rt}, where also the
scheme and scale dependences were estimated.
In Ref.~\cite{ref:Y}, 
the gauge-dependence problem of Ref.~\cite{ref:3} 
was remedied by adopting the pinch
technique. 
In Ref.~\cite{ref:4}, an alternative OS-like prescription was proposed that   
avoids this problem at one loop.   
The characteristic feature of this prescription is that the quark   
self-energies that enter the definition of $\delta V$ are not evaluated on   
their respective mass shells, but at the common subtraction point $q^2=0$.
To work exclusively in terms of OS renormalization constants, the authors of 
Ref.~\cite{ref:6} proposed to renormalize the CKM matrix with respect to a 
reference theory in which no quark mixing occurs.  
However, this prescription does not comply with the unitarity constraint for  
the renormalized CKM matrix, as was shown in Ref.~\cite{ref:7}, where this  
drawback was successfully eliminated.  
The renormalization prescriptions of Refs.~\cite{ref:Z,ref:DKR} are similar in 
spirit to the two-step procedure of Ref.~\cite{ref:7} and reach beyond the  
one-loop level. 
The prescription of Ref.~\cite{ref:L} is based on an {\it ad hoc} separation of
the one-loop quark self-energies into ultraviolet(UV)-divergent,
gauge-independent parts to be absorbed into the CKM matrix counterterm and
UV-finite, gauge-dependent parts to be combined with the vertex corrections. 
A genuine OS renormalization condition for the CKM matrix, which satisfies the 
criteria of UV finiteness, gauge independence, and unitarity has been
found recently \cite{ref:8}.  
It is based on a novel procedure to separate the external-leg mixing  
corrections into gauge-independent self-mass and gauge-dependent wave-function 
renormalization contributions.   
Very recently, a variant of the prescription of Ref.~\cite{ref:8} was proposed 
that is flavor democratic and formulated in terms of the invariant self-energy 
functions appearing in the quark mixing amplitudes \cite{ref:12}. 
The prescriptions of Refs.~\cite{ref:7,ref:8,ref:12} have the important
property that they are based on explicit OS renormalization conditions.   

The plan of this paper is the following.
In Sec.~\ref{sec:two}, we study numerically the effects of CKM matrix
renormalization on the hadronic partial decay widths of the $W$ boson at one
loop, on the basis of the CKM matrix elements $V_{ij}$ obtained in the global
analysis \cite{ref:10}.
For definiteness, we focus on the prescriptions of  
Refs.~\cite{ref:3,ref:4,ref:7,ref:8,ref:12}, which we also compare to the
modified minimal-subtraction ($\overline{\rm MS}$) scheme.  
We believe that these prescriptions are representative, since the others are 
either based on ideas similar to those in Refs.~\cite{ref:3,ref:4,ref:7}  
and/or do not comply with all the properties which the renormalized CKM quark  
mixing matrix should have, namely UV finiteness, gauge independence, 
and unitarity.  
Although the renormalization proposal of Ref.~\cite{ref:3} does not fulfill 
the second criterion, we include it in our analysis, as implemented in
't~Hooft-Feynman gauge, because it is the first attempt to renormalize the  
three-generation CKM matrix.
In Sec.~\ref{sec:three}, we use the results of Sec.~\ref{sec:two} to evaluate
the relative shifts in the $|V_{ij}|^2$ parameters induced by the
incorporation of the quark mixing renormalization effects.
This section contains also a discussion of the scheme dependence of these
shifts and their implications for the most precise unitarity test of the CKM
matrix, involving its first row.
Section~\ref{sec:four} summarizes our conclusions.

\section{Evaluation of the $W$-boson
hadronic widths}
\label{sec:two} 
   
We consider the two-particle decay of the $W^+$-boson to generic quarks,   
\begin{equation}   
W^+(k)\rightarrow u_i(p_1)\overline{d}_j(p_2).    
\label{process}    
\end{equation}   
The partial decay width in the Born approximation is given by   
\begin{eqnarray}   
\Gamma_0^{Wu_i\overline{d}_j}  
&=&\frac{N_c\alpha\left| V_{ij}\right|^2}{24s_w^2m_W^3}  
\kappa(m_W^2,m_{u,i}^2,m_{d,j}^2)  
\left[2m_W^2-m_{u,i}^2-m_{d,j}^2
\vphantom{\frac{(m_{u,i}^2-m_{d,j}^2)^2}{m_W^2}} 
\right.\nonumber\\ 
&&{}-\left.\frac{(m_{u,i}^2-m_{d,j}^2)^2}{m_W^2}\right],   
\end{eqnarray}   
where $N_c=3$, $\alpha=e^2/(4\pi)$ is the fine-structure constant, and   
\begin{equation}   
\kappa(x,y,z)=\sqrt{x^2+y^2+z^2-2(xy+yz+zx)}   
\end{equation}   
is K\"all\'en's function.   
   
The one-loop-corrected partial decay width is calculated by including the   
renormalization constants for the parameters $e$, $s_w$, and $V_{ij}$, those   
for the $W^+$, $u_i$, and $\overline{d}_j$ fields, and the proper vertex   
corrections.   
The results can be expressed in the form:   
\begin{equation}   
\Gamma_1^{Wu_i\overline{d}_j}=\Gamma_0^{Wu_i\overline{d}_j}   
(1+\delta^{\rm ew}+\delta^{\rm QCD}),   
\label{oneloopwidth}   
\end{equation}   
where $\delta^{\rm ew}$ and $\delta^{\rm QCD}$ are the electroweak and QCD   
corrections, respectively.   
Analytical expressions for $\delta^{\rm ew}$ and $\delta^{\rm QCD}$ in the  
$R_\xi$ gauges may be found, for example, in Ref.~\cite{ref:5}.   

Note that $\delta^{\rm ew}$ and $\delta^{\rm QCD}$ also receive contributions
from the bremsstrahlung of a single real photon and gluon, respectively.
Going beyond one loop, it is important to redefine the $W^+$-boson partial
decay widths so that they remain infrared-safe observables.
An obvious way of doing this is to generalize Eq.~(\ref{oneloopwidth}) to any
order beyond one loop by including all final-state configurations of the type
$u_i\overline{d}_j+X$, where $X$ comprises all possible sets of additional
particles, possibly including further $u_i$ or $\overline{d}_j$ quarks.
This represents a fully inclusive quantity, which is manifestly free of
infrared (soft and collinear) singularities by the Kinoshita-Lee-Nauenberg
theorem.
This definition also avoids the use of jet algorithms and fragmentation
functions altogether, which could dilute the sensitivity to the CKM matrix
elements.
    
We now proceed with our numerical analysis of Eq.~(\ref{oneloopwidth}).   
We perform all the calculations with the aid of the {\tt LOOPTOOLS}  
\cite{ref:9} package embedded into the {\tt MATHEMATICA} environment.   
As a check, we reproduce the numerical results of Ref.~\cite{ref:5} when   
adopting the definition of $\delta V_{ij}$ and the values of the input   
parameters employed in that paper.  
   
In our analysis, we use the following input parameters \cite{ref:10}:   
   
\vspace*{1cm}   
\begin{tabular}[b]{l@{\vrule height 12pt depth 4pt width 0pt\hskip\arraycolsep}ll}   
$\alpha=1/137.035999679$, & $G_F=1.16637\times 10^{-5}~{\rm GeV}^{-2}$,  
& $\alpha_s^{(5)}(m_Z)=0.1176$, \\   
$m_W=80.398~{\rm GeV}$, & $m_Z=91.1876~{\rm GeV}$, &  \\   
$m_e=0.510998910~{\rm MeV}$,\ \ & $m_\mu=105.658367~{\rm MeV}$,   
& $m_\tau=1776.84~{\rm MeV}$, \\   
$m_u=2.4~{\rm MeV}$, & $m_d=4.8~{\rm MeV}$, & $m_s=100~{\rm MeV}$, \\   
$m_c=1.25~{\rm GeV}$, & $m_b=4.25~{\rm GeV}$, & $m_t=172.4~{\rm GeV}$.   
\end{tabular}   
\vspace*{1cm}   
   
The standard parameterization of the CKM matrix, in terms of the three mixing
angles $\theta_{ij}$ and the CP-violating phase $\delta$, reads \cite{ref:10}: 
\begin{equation}   
 V=\left(\begin{array}{ccc}   
c_{12}c_{13} & s_{12}c_{13} & s_{13}e^{-i\delta} \\  
-s_{12}c_{23}-c_{12}s_{23}s_{13}e^{i\delta}  
& c_{12}c_{23}-s_{12}s_{23}s_{13}e^{i\delta} & s_{23}c_{13} \\  
s_{12}s_{23}-c_{12}c_{23}s_{13}e^{i\delta}  
& -c_{12}s_{23}-s_{12}c_{23}s_{13}e^{i\delta} & c_{23}c_{13}  
\end{array}\right)   
\label{ckmdef1}   
\end{equation}   
where $s_{ij}=\sin\theta_{ij}$ and $c_{ij}=\cos\theta_{ij}$.  
The choice   
\begin{equation}   
s_{12}=\lambda,\qquad  
s_{23}=A\lambda^2,\qquad  
s_{13}e^{i\delta}=\frac{A\lambda^3(\overline{\rho}+i\overline{\eta})  
\sqrt{1-A^2\lambda^4}}  
{\sqrt{1-\lambda^2}[1-A^2\lambda^4(\overline{\rho}+i\overline{\eta})]}   
\label{ckmdef2}   
\end{equation}   
ensures that the CKM matrix written in terms of $\lambda$, $A$,  
$\overline{\rho}$, and $\overline{\eta}$ is unitary to all orders in  
$\lambda$.   
In our analysis, we evaluate the CKM matrix elements from Eqs.~(\ref{ckmdef1}) 
and (\ref{ckmdef2}) using the values $\lambda=0.2257$, $A=0.814$,   
$\overline{\rho}=0.135$, and $\overline{\eta}=0.349$ \cite{ref:10}.   

\section{Results} 
\label{sec:three} 
   

\begin{table}[t]   
\centering   
{\footnotesize{   
\begin{tabular}[b]{|l@{\vrule height 12pt depth 4pt width 0pt  
\hskip\arraycolsep}|c|c|c|c|c|c|c|c|}\hline\hline   
Partial width & Ref.~\cite{ref:3} & Ref.~\cite{ref:4} & Ref.~\cite{ref:7} &  
Ref.~\cite{ref:8} & Ref.~\cite{ref:12} & ${\overline {\rm MS}}$ scheme &
$\delta V_{ij}=0$ \\  
\hline   
$\Gamma(W^+\to u\overline{d})$ & 0.6697016 & 0.6697016 & 0.6697016 & 0.6697016  
& 0.6697016 &0.6696999 & 0.6697012 \\   
$\Gamma(W^+\to u\overline{s})\times 10$ & 0.3594604 & 0.3594604 & 0.3594604  
& 0.3594604 & 0.3594604 & 0.3594804 & 0.3590518 \\   
$\Gamma(W^+\to u\overline{b})\times 10^5$ & 0.9345792 & 0.9309188 & 0.9345444  
& 0.9345781 & 0.9345797 & 0.9040684 & 0.9065685 \\   
$\Gamma(W^+\to c\overline{d})\times 10$ & 0.3589746 & 0.3589746 & 0.3589746   
& 0.3589746 & 0.3589746 & 0.3589738 & 0.3556135 \\   
$\Gamma(W^+\to c\overline{s})$ & 0.6684818 & 0.6684818 & 0.6684819 & 0.6684818  
& 0.6684818 & 0.6684267 & 0.6614634 \\   
$\Gamma(W^+\to c\overline{b})\times 10^2$ & 0.1211309 & 0.1211315 & 0.1211263  
& 0.1211318 & 0.1211315 & 0.1266316 & 0.1196919 \\\hline   
$\Gamma(W\to{\rm hadrons})$ & 1.4112476 & 1.4112476 & 1.4112476 & 1.4112476  
& 1.4112476 & 1.4112474 & 1.4038372 \\\hline\hline   
\end{tabular}}}   
\caption{Partial widths (in GeV) of the hadronic $W$-boson decay channels
evaluated at one loop using the quark mixing renormalization prescriptions of
Refs.~\cite{ref:3,ref:4,ref:7,ref:8,ref:12} and the ${\overline{\rm MS}}$
scheme.
The entries of the last column are obtained by neglecting quark mixing
renormalization.\label{tab1}}   
\end{table}     
In Table~\ref{tab1}, the one-loop-corrected partial widths of the various 
hadronic $W$-boson decay channels are presented for the selected definitions 
of the CKM counterterm matrix $\delta V_{ij}$ 
\cite{ref:3,ref:4,ref:7,ref:8,ref:12}, assuming $m_H=120$~GeV.   
The first and second columns in Table~\ref{tab1} (not counting the one labeled
{\it Partial width}) describe the partial widths of the $W$ boson when adopting
the CKM matrix renormalization conditions proposed in Refs.~\cite{ref:3,ref:4},
respectively.   
This has been already done in the literature, for example in   
Ref.~\cite{ref:5}.   
We emphasize that we find full agreement, provided we adopt the same values   
for the input parameters.   
Note that the prescription of Ref.~\cite{ref:3} leads to a gauge-dependent   
result, so that the gauge choice must be specified.   
We perform the calculation in 't~Hooft-Feynman gauge.   
   
New results are those from the third, fourth, and fifth columns, which refer to
the three genuine OS renormalization proposals of
Refs.~\cite{ref:7,ref:8,ref:12}, respectively.  
The prescription of Ref.~\cite{ref:7} entails the minor complication that one  
needs to consider a reference theory with zero quark mixing.  
It is important to note that the proposals of Refs.~\cite{ref:7,ref:8,ref:12}  
have the important property that they lead to renormalized amplitudes that are
non-singular in the limit in which any two fermions become mass degenerate and
are thus suitable for the generalization to theories where maximal mixing  
could appear.  
A generalization of Ref.~\cite{ref:8} to lepton mixing in Majorana-neutrino  
theories has recently been carried out in Ref.~\cite{ref:11}.  
For reference, we have included in the sixth column of Table~\ref{tab1} the  
results based on the $\overline{\rm MS}$ scheme with 't~Hooft mass scale  
$\mu=m_W$.
Finally, in order to assess the significance of quark mixing renormalization,
we have included in the last column the results of 
calculations where $V_{ij}=\delta_{ij}$ is substituted in loops inserted in
the external quark legs, so that the criteria of UV finiteness,
gauge independence, and unitarity may be satisfied with the trivial choice
$\delta V_{ij}=0$.
This corresponds to the conventional calculations in which mixing effects in
the external quark legs are neglected, on the grounds that their UV
divergences are canceled by the counterterms and their finite contributions
are very small.
The numbers in Table~\ref{tab1} are not meant to give the $W$-boson decay 
widths with the stated accuracy, since they are based on a one-loop 
calculation. 
However, it is necessary to exhibit 7 digits in the one-loop results in order
to illustrate their differences.

It is important to note that these corrections also affect the theoretical
calculations of the accurate observables underpinning the determination of the
CKM elements $V_{ij}$.
When inserted into those calculations, they lead to modified parameters
$|V_{ij}^\prime|^2$ that cancel, at the one-loop electroweak level, the very
small scheme dependence portrayed in Table~\ref{tab1}.
In order to show this cancellation, we call $\delta_{ij}^\alpha$ the one-loop
correction in renormalization scheme $\alpha$, and $\delta_{ij}^0$ the one
corresponding to the last column in Table~\ref{tab1}.
Taking into account that in the conventional determination of the $V_{ij}$
parameters quark mixing effects in the external legs are generally neglected,
as is also the case in $\delta_{ij}^0$, we readily find the relation:
\begin{equation}
|V_{ij}^{\prime\alpha}|^2(1+\delta_{ij}^{\alpha})
=|V_{ij}|^2(1+\delta_{ij}^0),
\end{equation}
where $\alpha$ labels the renormalization scheme employed.
In turn, this implies
\begin{equation}
\frac{|V_{ij}^{\prime\alpha}|^2}{|V_{ij}|^2}=R_{ij}^\alpha,
\label{eq:R}
\end{equation}
where $R_{ij}^\alpha$ are the ratios of the entries in the last column in
Table~\ref{tab1} and those in the $\alpha$ column.
In order to incorporate the modified CKM parameters in the calculation of the
partial widths, we multiply the entries in the first six columns of that Table
by $|V_{ij}^{\prime\alpha}|^2/|V_{ij}|^2$ and, using Eq.~(\ref{eq:R}), we see
that they become equal to those in the last column, independently of the chosen
renormalization scheme $\alpha$.
In summary, when the calculations of the $W$-boson decay widths incorporate the
modified CKM parameters $|V_{ij}^{\prime\alpha}|^2$, the very small scheme
dependence portrayed in Table~\ref{tab1} cancels.

\begin{table}[t]
\centering   
{\footnotesize{   
\begin{tabular}[b]{|c@{\vrule height 12pt depth 4pt width 0pt  
\hskip\arraycolsep}|c|c|c|c|c|c|c|c|}\hline\hline   
$\Delta_{ij}^\alpha$ & Ref.~\cite{ref:3} & Ref.~\cite{ref:4} & 
Ref.~\cite{ref:7} &  Ref.~\cite{ref:8} & Ref.~\cite{ref:12} & 
${\overline {\rm MS}}$ scheme & $|V_{ij}|^2$ \cite{ref:10} \\  
\hline
$ud$ & $-5.29\times10^{-5}$ & $-5.29\times10^{-5}$ & 
$-5.26\times10^{-5}$ & $-5.29\times10^{-5}$ & $-5.29\times10^{-5}$ &
$2.00\times10^{-4}$ & 0.94905 \\   
$us$ & $-0.114$ & $-0.114$ & $-0.114$ & $-0.114$ & $-0.114$ & 
$-0.119$ & $5.0940\times10^{-2}$ \\   
$ub$ & $-3.00$ & $-2.62$ & $-2.99$ &  $-3.00$ & $-3.00$ & $0.277$ &
$1.2888\times10^{-5}$ \\   
$cd$ & $-0.936$ & $-0.936$ & $-0.936$ & $-0.936$ & $-0.936$ &
$-0.936$ & $5.0895\times10^{-2}$ \\   
$cs$ & $-1.05$ & $-1.05$ & $-1.05$ & $-1.05$ & $-1.05$ & $-1.04$ &
0.94739 \\   
$cb$ & $-1.19$ & $-1.19$ & $-1.18$ & $-1.19$ &  $-1.19$ &
$-5.48$ & $1.7223\times10^{-3}$ \\
\hline\hline   
\end{tabular}}}   
\caption{Relative shifts $\Delta_{ij}^\alpha$ (in \%) in the central values of
$|V_{ij}|^2$ \cite{ref:10} induced by quark mixing renormalization effects
according to the prescriptions $\alpha$ of
Refs.~\cite{ref:3,ref:4,ref:7,ref:8,ref:12} and the ${\overline {\rm MS}}$
scheme.\label{tab2}}   
\end{table}     
On the other hand, Eq.~(\ref{eq:R}) permits us to evaluate the relative shifts,
\begin{equation}
\Delta_{ij}^\alpha
=\frac{|V_{ij}^{\prime\alpha}|^2-|V_{ij}|^2}{|V_{ij}|^2}=R_{ij}^\alpha-1,
\end{equation}
in the $|V_{ij}|^2$ parameters induced by the quark mixing renormalization
effects, an issue of considerable interest given the fundamental importance
of the CKM parameters.
The results are portrayed in Table~\ref{tab2}.
(In order to compute some of the entries in Table~\ref{tab2}, we have used
more precise values than those displayed in Table~\ref{tab1}.)

From Table~\ref{tab2} we see that the scheme dependence of
$\Delta_{ij}^\alpha$ among the five prescriptions
\cite{ref:3,ref:4,ref:7,ref:8,ref:12} is extremely small, of
${\cal O}(10^{-2}\%)$ or less, except in the single case of
$|V_{ub}^\prime|^2$ in scheme \cite{ref:4}, where it reaches 0.38\%.
The differences in the $\Delta_{ij}^\alpha$ between those five schemes and the
${\overline {\rm MS}}$ evaluations are also very small, of
${\cal O}(10^{-2}\%)$ or less, except in $|V_{ub}^\prime|^2$ and
$|V_{cb}^\prime|^2$, where they reach 3.3\% and 4.3\%, respectively.

A matter of considerable interest is the magnitude of $\Delta_{ij}^\alpha$.
With only two exceptions in the ${\overline {\rm MS}}$ scheme, a general
feature is that the incorporation of the quark mixing renormalization effects
decreases the values of the $|V_{ij}|^2$ parameters.
In particular, using the results in the first five columns of
Table~\ref{tab2}, we see that $|V_{ud}|^2$ is not modified to a high degree of
accuracy, $|V_{us}|^2$ is decreased by 0.11\%, $|V_{ub}|^2$ by 3.0\%,
$|V_{cd}|^2$ by 0.94\%, $|V_{cs}|^2$ by 1.1\%, and $|V_{cb}|^2$ by 1.2\%.

We now consider the effect of these shifts on the most precise unitarity test
of the CKM matrix, involving the elements in its first row.
The latest update \cite{bill} employs $|V_{ud}|=0.97425(23)$ and
$|V_{us}|=0.2252(9)$, values that differ slightly from those reported in
Ref.~\cite{ref:10}.
They lead to
\begin{equation}
|V_{ud}|^2+|V_{us}|^2+|V_{ub}|^2=0.9999(6),
\end{equation}
in excellent agreement with unitarity.

Including the quark mixing renormalization effects discussed in this paper, we
have $|V_{ud}^\prime|=0.97425(23)$, since $|V_{ud}|$ is not altered,
$|V_{us}^\prime|=0.2251(9)$, and $|V_{ub}^\prime|=0.00354(16)$, leading to
\begin{equation}
|V_{ud}^\prime|^2+|V_{us}^\prime|^2+|V_{ub}^\prime|^2=0.9998(6).
\label{eq:uni}
\end{equation}
We note that the shifts in the $|V_{ij}|$ parameters and the unitarity test
are considerable smaller than the current errors in their evaluation.
On the other hand, Eq.~(\ref{eq:uni}) remains an impressive test of the SM at
the level of its quantum corrections!
In fact, it is worth remembering that the electroweak corrections in this test
amount to roughly 4\% \cite{Sirlin:1977sv}.
Thus, if they were neglected, the unitarity test of the CKM matrix would fail
by about 66 standard deviations!
   
\section{Conclusions} 
\label{sec:four} 
 
In summary, we have reviewed a number of schemes for the renormalization of
the CKM matrix and studied the numerical effects of several of them on the
$W$-boson hadronic partial decay widths, using the $V_{ij}$ values obtained in
the global analysis.
We have then employed these results to infer the relative shifts in the
$|V_{ij}|^2$ parameters due to the quark mixing renormalization corrections.
Finally, we have discussed the effect of these shifts on the most precise
unitarity test of the CKM matrix.


\begin{acknowledgments}
B.A.K. and A.S. are grateful to the Max Planck Institute for Physics in Munich 
for the warm hospitality during a stay when part of this work was carried out. 
A.S. thanks the Institute for Nuclear Theory at the University of Washington
for its hospitality and the Department of Energy for partial support during
the completion of this work.
This work was supported in part by the German Research Foundation through the 
Collaborative Research Center No.~676 {\it Particles, Strings and the Early
Universe --- the Structure of Matter and Space Time}.  
The work of A. Sirlin was supported in part by the National Science Foundation 
through Grant Nos.\ PHY--0245068 and PHY--0758032. 
\end{acknowledgments}

\end{document}